\documentstyle[eqsecnum,aps,epsf,rotate,preprint]{revtex}
\begin{document}
\draft
\input psfig.tex

\title{Self--organized criticality due to a separation of energy scales}
\author{Barbara Drossel}
\address{Department of Physics, Massachusetts Institute of  
Technology, Cambridge, Massachusetts 02139}
\address{Tel.: (617) 253-6133; Fax: (617) 253-2562; email: drossel@cmt0.mit.edu}
\date{\today}
\maketitle
\begin{abstract}
Certain systems with slow driving and avalanche--like dissipation events are naturally close to a critical point when the ratio of two energy scales is large. The first energy scale is the threshold above which an avalanche is triggered, the second scale is the threshold above which a site is affected by an avalanche. I present results of computer simulations, and a mean--field theory. 
\end{abstract}
\pacs{PACS numbers: 05.40.+j, 05.70.Jk, 05.70.Ln. \hfill\break
}

\section{Introduction}
\label{u1}
During the past years, systems  which exhibit self--organized criticality (SOC)
have attracted much attention, since they might explain part of the abundance
of fractal structures in nature  \cite{bak87}. Their common
features are slow driving or energy input (e.g. dropping of sand grains \cite{bak87},
increase of strain \cite{ola92}, tree growth \cite{dro92}, spontaneous mutations \cite{sne93}) and rare dissipation events which are
instantaneous on the time scale of driving (e.g. sand avalanches, earthquakes, fires, or a series of rapid mutations).  
In the stationary state, the size distribution of dissipation events 
obeys a power law, irrespective of initial conditions and without the need to
fine-tune parameters. 
There is, however, no reason to expect that systems with slow driving and instantaneous avalanches are always SOC. Such 
systems might also have many small avalanches which release only little energy,
or only large avalanches which release a finite part of the system's energy, 
or some combination of both. SOC systems are naturally at the 
critical point, due to a conservation law (sandpile model), a second time scale separation (forest--fire model),  a competition between open boundary conditions and the tendency of 
neighboring sites to synchronize (earthquake model \cite{ola92,mid95}), or due to the slow driving alone ( ``evolution'' model). Often, the critical behavior breaks down when details of the model rules are changed (e.g. the boundary conditions in the earthquake model \cite{mid95} or the tree growth rule in the forest--fire model \cite{dro96}).  

There are certain parallels between these models and equilibrium critical systems, since both consist of many small units which interact with their neighbors, and since spin clusters in an Ising model or clusters of occupied sites in percolation theory can be compared to avalanches. However, the critical behavior of nonequilibrium systems can depend on microscopic details, as mentioned above, in contrast to equilibrium critical phenomena, which commonly show universal behavior. Also, nonequilibrium systems do not satisfy a detailed-balance condition and can e.g.  show periodic behavior. Furthermore, avalanches are usually released when some variable reaches locally a threshold, while other regions of the system might be far below the threshold,  and consequently not all parts of the system look equal. This can in particular result in more than one diverging length scale, as in the earthquake model \cite{mid95} or in the forest--fire model \cite{cla95,hon96}. By contrast, in equilibrium systems the energy is an extensive variable, which means that all regions (which are large compared to the lattice size) are equal. This is e.g. the basis for hyperscaling relations. 

In this paper, I will add to the rich scenario of SOC systems by discussing models with slow driving and avalanche--like dynamics that show SOC behavior of a new kind. They have two threshold energies $E_c$ and $E_0$ and become critical with a power-law size distribution of avalanches in the limit of a separation of {\it energy scales}, $(E_c - E_0)/E_0 \to \infty$. The size of the largest avalanche (i.e. the cutoff in the power law) depends on the degree of energy scale separation. These conclusions are derived from computer simulations in $d=2$ dimensions, from intuitive arguments, and from analytical mean--field calculations. I will also argue that the critical behavior breaks down when certain details of the rules are changed, leading to synchronized dynamics. 

The outline of this paper is as follows: In section \ref{u2}, the basic model is defined, and simulation results in two dimensions are presented. Section \ref{u3} defines related models that have the same behavior as shown by simulation results.  Section \ref{u4} discusses the mean--field theory, and section \ref{u5} contains the conclusions.

\section{The model}
\label{u2}
The model presented in this section is inspired by a deterministic forest--fire model introduced in \cite{che90} a few years ago as a toy model for turbulence. Since that paper discusses the model only for one choice of the parameters, the generalization given below is natural and provides a deeper understanding of the model. The model is defined on a $d$--dimensional hypercubic lattice with $L^d$ sites. In the beginning, each site is assigned a variable $E$ which is randomly chosen from the interval $[0,E_c)$ and can be interpreted as ``tree height'' or ``energy''. 
Sites with a height between $0$ and $E_0$ are called ``empty'', sites with a height between $E_0$ and $E_c$ are called ``tree'', and sites with a height above $E_c$ are called ``burning''. All heights are increased globally and very slowly. After some time, one of them reaches the threshold $E_c$ and becomes a burning tree. Fire spreads to trees on the $2d$ neighboring sites, and burning trees loose part of their height. After a fire has died, all heights increase until another tree reaches the threshold height, releasing a new fire.  Driving is so slow that only a negligeable amount of energy is put into the system while a fire is spreading. In the simulations, the system is not driven at all during a fire, and at the end of a fire the energy of all sites is increased by the difference between $E_c$ and the largest height in the system. The precise rules for the spreading of the fire are the following:
\begin{equation}
E \longrightarrow E + {E_c - E_0}, \text{ if a nearest neighbor is burning and $E \in \left[E_0,E_c\right)$} \,. \label{rule1}
\end{equation}
\begin{equation}
E \longrightarrow (E - E_c)\, {E_c \over E_c - E_0},  \text{ if } E \ge E_c\,. \label{rule2}
\end{equation}
 They are applied to each site at each time step, as long as there exist burning sites.
The net effect of a fire on a tree  of height $E \in \left[E_0,E_c\right)$ is a reduction in height to the new value $(E - E_0) E_c/(E_c - E_0)$. The interval $\left[E_0,E_c\right)$ is linearly mapped on the interval $\left[0,E_c\right)$. The distance in energy between neighboring sites therefore increases when both sites are burnt, making any kind of synchronization impossible for this model, and leading to chaotic behavior. 
A site can burn several times during a fire, until its height or the height of all its nearest neighbors is below $E_0$. We define the size $s$ of a fire as the number of sites burnt during a fire, including multiple burnings. 

After a certain time, the systems reaches a stationary state with a certain mean energy and a certain size distribution of fires (averaged over either an ensemble of systems or many time steps). In this paper, we are only interested in the properties of this stationary state, and not in the transient behavior of the first time steps. Since the model is deterministic, chaotic, and dissipative, the stationary state is a strange attractor which occupies a vanishing fraction of the $L^d$--dimensional phase space (however, due to rounding errors in the computer, the model is not completely deterministic). Since the dynamics of the system are invariant when all energies (including the thresholds) are multiplied by the same factor, variation of $E_c$ alone gives all information about the system, and we can limit the discussion to the case $E_0=1$.
In the original version of the model, $E_0=1$ and $E_c = 2$ \cite{che90}, and the size distribution of fires is a power law over a few orders of magnitude \cite{soc93}. 

\begin{figure}
\centerline{{\epsfysize=4in 
\epsffile{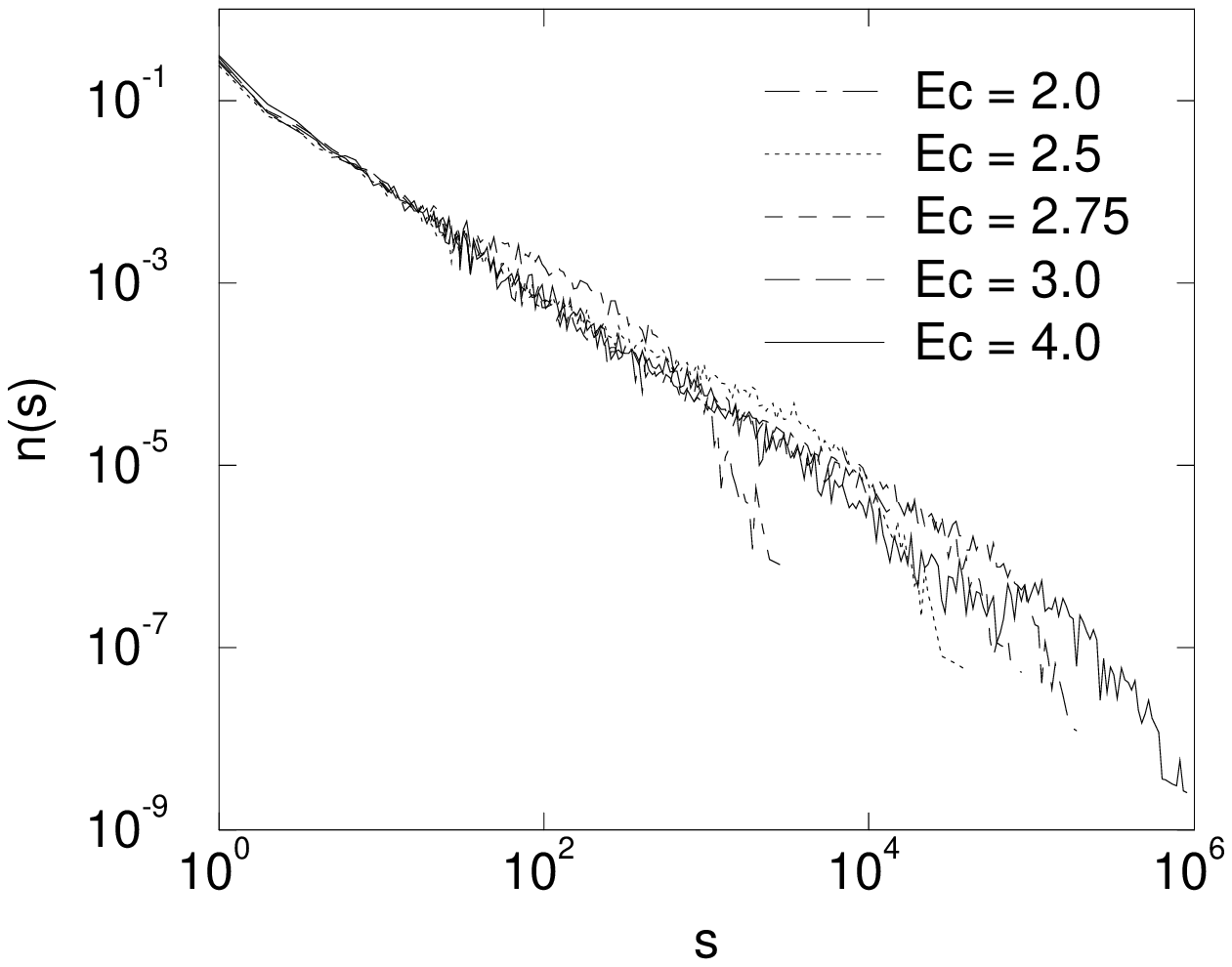}}}
\vskip -1cm
\caption{Size distribution of fires for $L=400$ in the deterministic model.}
\label{deta}
\end{figure}

\begin{figure}
\centerline{{\epsfysize=4in 
\epsffile{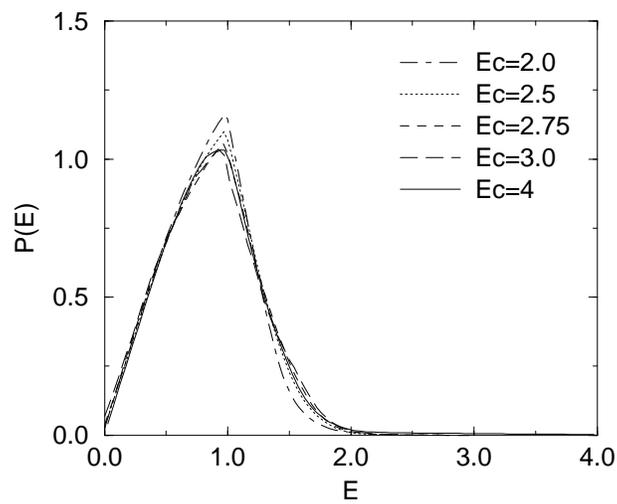}}}
\vskip -1cm
\caption{Probability distribution of energies for $L=400$ in the deterministic model.}

\label{detb}
\end{figure}

When $E_c$ is equal to $E_0 \equiv 1$, there are only fires of size one, since all neighbors of a site reaching the threshold are still empty. When $E_c$ is slightly above $E_0$, sometimes neighbors of a tree reaching the threshold are between 1 and $E_c$, and fires burning several sites can occur. There is a cutoff in fire size which depends on $E_c$. With increasing $E_c$, the ratio between the number of trees and empty sites becomes larger, and the cutoff in fire size increases. We expect therefore fires spreading over the whole system when $E_c$ is larger than some critical value $E_c^*$. 
The simulation results, however, do not show such a transition for any value of $E_c$ that can be studied with linear system sizes $L \leq 400$. Fig.~\ref{deta} shows the size distribution $n(s)$ of fires for $L=400$. The simulations were started with a random energy distribution. To assure that the system is in a stationary state, the first 10000 simulation steps were discarded, and $n(s)$ was determined from the following 10000 steps. The fire size distribution decays as $s^{-\tau}$ with $\tau \simeq 1.07 \pm 0.03$, and with a cutoff for larger fire sizes. Simulations for $E_c > 3.0$ show finite--size effects since they lead to avalanches of the order of the system size $L^2$. This explains the bump in the fire size distribution for $E_c = 4.0$. Similiar bumps occur already for smaller values of $E_c$ when the system size is smaller. Of course, such bumps should also occur when the system is beyond a critical point and has true infinite avalanches. But the simulations do not show the other signatures expected for a phase transition: Beyond a critical point, the bump size increases with increasing distance from the critical point, and the cutoff in cluster size diverges when $E_c^*$ is approached from below. 
Although it is impossible to derive from our simulation results the analytical  dependence of the cutoff cluster size on $E_c$, we can clearly see that it has no tendency to diverge when $E_c$ increases. Although these simulation results allow no conclusive prediction whether there exists a finite $E_c^*$ that is much larger than the values studied in the simulations, it seems well possible that the critical energy is infinity. This hypothesis is supported by the probability distribution of energy in the system, shown in Fig.~\ref{detb}. This probability distribution has a rapid decay for energies larger than $E_0 \equiv 1$,
and seems to approach a constant form as $E_c$ increases. 
This means that the fraction of excitable sites is almost the same for all energies, indicating that there is no phase transition in the system. 
A simple consideration shows that the mean fire size $\bar s = \sum_{s=1}^\infty sn(s)$ diverges in the limit $E_c \to \infty$ if the function $P(E)$ approaches a constant form: Let $v$ be the (very slow) driving velocity. Per unit time $\hbox{d}t$, $L^2 \bar s P(E_c) v\hbox{d}t$ sites participate in a fire. If the function $P(E)$ approaches a fixed function in the limit  $E_c \to \infty$, this number must also be independent of $E_c$. Consequently, $\bar s P(E_c)$ must approach a constant for  $E_c \to \infty$. If $P(E)$ changes only slightly with increasing $E_c$, only a vanishing fraction of sites can have large energies, and $P(E_c) \to 0$ for $E_c \to \infty$. This implies that $\bar s \to \infty$. 

The following intuitive argument gives a deeper understanding why $E_c^*$ cannot be finite in our model: Assume that there is a finite $E_c^*$, above which the mean fire size becomes proportional to the system size, $\bar s \propto L^2$. These infinite fires dissipate a finite fraction of the system's energy. Consequently, a finite fraction of the system's energy must be put into the system between fires. This is only possible if the highest energy in the system after an fire is at a finite distance from $E_c$. However, the rules of our model do not allow such an energy gap, since the energy of sites that are very close to $E_c$ is decreased only by a very small amout if this site is involved in a fire. Since this argument is very general, we expect a similiar behavior also in higher dimensions. In the language of critical phenomena this means that the lower critical dimension of the model is infinity. 

The scenario in one dimension is slightly different. Since any site with an energy below $E_0\equiv 1$ stops the fire, less and less sites have energies below $E_0$ as $E_c$ increases and the mean fire size becomes larger. The function $P(E)$ therefore does not approach any finite limit function as $E_c$ diverges. 

Once we have understood the mechanism leading to $E_c^* = \infty$, we can easily construct other models with the same behavior. The next section introduces two of these models and shows simulation results in two dimensions. Section \ref{u4} then presents the mean--field theory, showing explicitely that $E_c^* = \infty$. Since the mean--field theory usually describes well systems in high dimensions, this analytical calculation supports the conclusion that the behavior observed in two-dimensional simulations persists in higher dimensions. 

\section{Related models}
\label{u3}
\begin{figure}
\centerline{{\epsfysize=4in 
\epsffile{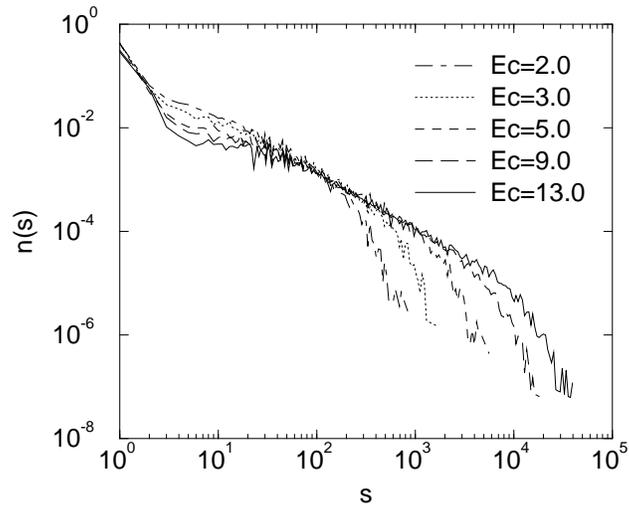}}}
\vskip -1cm
\caption{Size distribution of fires for $L=400$ in the first stochastic model.}
\label{stoc1a}
\end{figure}

\begin{figure}
\centerline{{\epsfysize=4in 
\epsffile{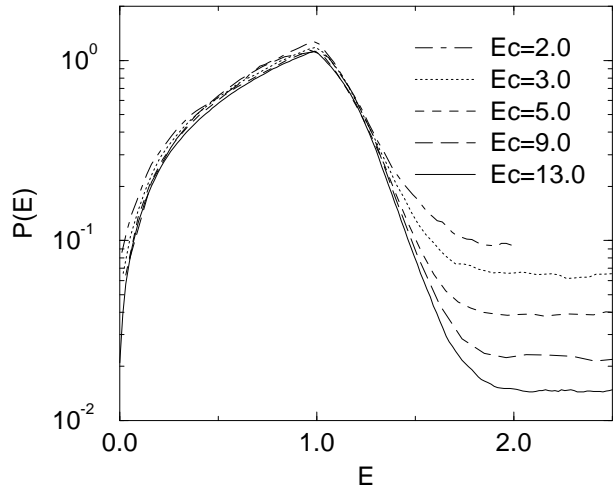}}}
\caption{Probability distribution of energies for $L=400$ in the first stochastic model.}
\label{stoc1b}
\end{figure}

\begin{figure}
\centerline{{\epsfysize=4in 
\epsffile{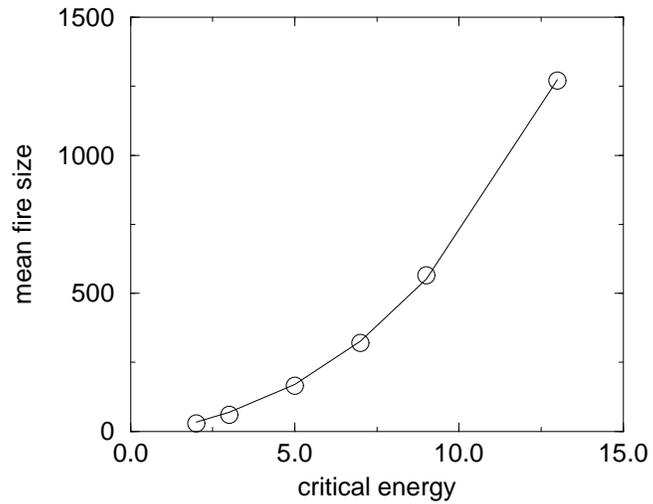}}}
\vskip -1cm
\caption{Mean fire size $\bar s$ as fuction of the critical energy $E_c$. The solid 
line is a stretched exponential $\bar s \propto \exp[-2.4(E_c-1)^{0.37}]$.}
\label{stoc1c}
\end{figure}

\begin{figure}
\centerline{{\epsfysize=4in 
\epsffile{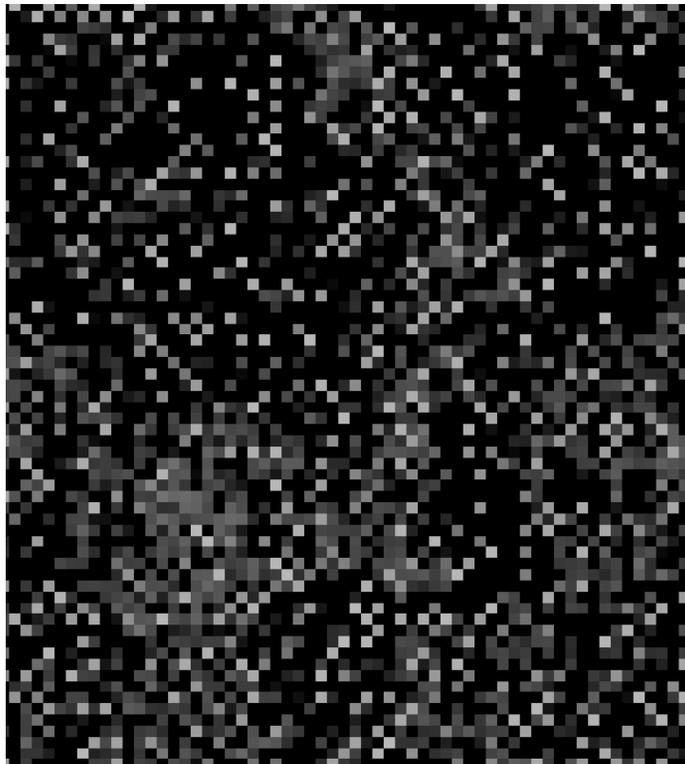}}}
\caption{Shapshot of the first stochastic model. Sites with $E<E_0$ are black; the energy of the other sites is indicated by the grey shade.}
\label{stoc1d}
\end{figure}

The above model is difficult to investigate analytically, since the energy of a site after a fire is related to the energy before the fire in a  nontrivial way, leading to complicated correlations. Models where the energy after a fire is essentially determined by a random--number generator are easier to understand. Let us consider therefore first a model that has the following version of rule 2 (Eq.~(\ref{rule2})) above:
\begin{equation}
E \longrightarrow E_c \times rand \,, \text{ if } E \ge E_c\,. \label{rule2neu}
\end{equation}
$rand$ is a random number equally distributed in the interval $[0,1)$. Like the previous model, this model does not allow an energy gap between the highest energy and $E_c$, and we expect again $E_c^* = \infty$. Since the fire jumps back and forth between neighboring sites until at least one of them is below $E_0\equiv 1$, a site that has participated in a fire is below $E_0$ with a probability larger than 0.5 even for large $E_c$, suggesting again a fixed limit $\lim_{E_c \to \infty} P(E)$, with only very few sites with high energies. Since all energies between zero and $E_0$ are chosen with equal probability according to rule (\ref{rule2neu}), we expect that $P(E)$ is linear in the interval $[0,E_0]$, as can also be seen from the explicit calculations in the next section. I performed the simulations under similar conditions as for the previous model. The resulting distributions $n(s)$ and $P(E)$ are shown in figures \ref{stoc1a} and \ref{stoc1b}. The exponent $\tau$ characterizing the fire size distribution is $\tau = 1.08 \pm 0.02$. This value is compatible with the one obtained for the deterministic model, and the two models possibly belong to the same universality class. The probability distribution of energies $P(E)$ is linear for $E<E_0$ (this cannot easily be seen in this logarithmic plot), and seems then to dacay exponentially, reaching a constant value proportional to $1/(E_c-E_0)$, a behavior that will also be obtained from the mean--field theory below. In order to make the interesting part of the curves more visible, the horizontal part has been cut off for energies larger than 3. Fig.~\ref{stoc1c} shows the mean fire size $\bar s$ as function of $E_c$. The best fit is obtained from a stretched exponential $\bar s \propto \exp[c(E_c-E_0)^\gamma]$, with $c=2.4$ and $\gamma=0.37$. We will see below that the mean--field theory gives a simple exponential increase. Fig.~\ref{stoc1d} shows a snapshot of part of the system for $E_c=5.0$. All sites with $E<1$ are black, other sites have a grey shade that indicates the energy. The scale for the grey shade is not linear.

\begin{figure}
\centerline{{\epsfysize=4in 
\epsffile{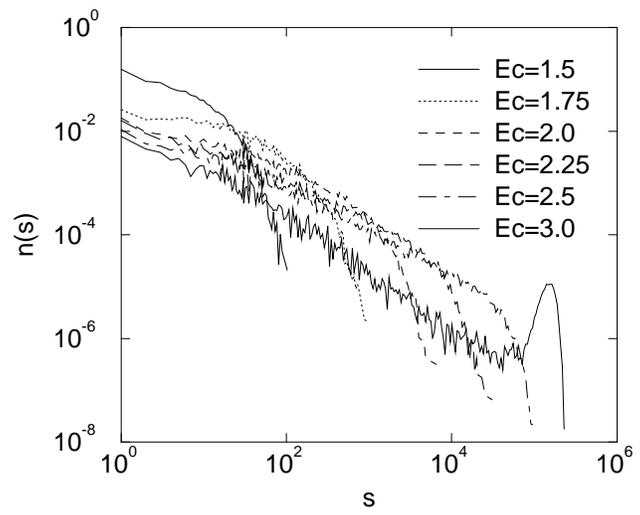}}}
\vskip -1cm
\caption{Size distribution of fires for $L=400$ in the second stochastic model.}
\label{stoc2}
\end{figure}

A model that is closer to the deterministic model of the previous section and that can also be analyzed in mean--field theory has the following rule 2:
\begin{equation}
E \longrightarrow (E-E_c+E_0) \times rand \,, \text{ if } E \ge E_c\,. \label{rule23}
\end{equation}
With this rule, the energy of a site is always decreased during a fire. Nevertheless, the probability that a given site will have an energy close to $E_c$ after a fire does not vanish, and we expect again $E_c^* = \infty$. 
The fire size distribution for this model is shown in figure \ref{stoc2}. The cutoff fire size increases much faster than in the previous model, and we cannot perform simulations with $E_c > 2.5$ without getting finite-size effects. This is because large energies occur now with a much smaller probability than before, reducing $P(E_c)$. If we assume that the fraction of sites below the threshold $E_0\equiv 1$ is of the same order in different models, it follows immediately that $\bar s$ (which is proportional to $1/P(E_c)$ as we have seen before) increases. Unfortunately, we cannot even extract a useful information about $\tau$ from the simulation results, since the system is still far from the asymptotic behavior.

\section{mean--field theory}
\label{u4}
The two models defined in the previous section can be solved analytically in mean--field theory. In a mean--field theory, all correlations between different sites are neglected, and the density $P(E)$ is calculated self-consistently. A random--neighbor version of the above models shows the same behavior as the mean--field theory. In this version, the $z=2d$ neighbors of a burning site are chosen at random for each simulation step. In the original models, the fire jumps back and forth between  neighboring sites, until at least one of the two has an energy below $E_0$. We take this into account in the mean--field theory by requiring that a finite fraction $\alpha$ of all sites that participate in a fire end up with an energy below $E_0$, where $\alpha$ is independent of $E_c$. The mean--field equations for the first stochastic model that assigns a random energy from the interval $[0,E_c)$ to each burning tree are the following
\begin{eqnarray}
{\partial P(E,t) \over \partial t} &=& - v {\partial P(E,t) \over \partial E} + v\alpha \bar s P(E_c)/E_0 \, \hbox{ for } E \le E_0\equiv 1 \, ;\nonumber \\
{\partial P(E,t) \over \partial t} &=& - v {\partial P(E,t) \over \partial E} + (1-\alpha){v \bar s P(E_c,t)\over E_c - E_0} - {v P(E,t) P(E_c,t)(\bar s - 1)\over \int_{E_0}^{E_c} P(E',t)\hbox{d}E'} \,\nonumber \\
&& \hskip 2cm \hbox{ for } E_0<E<E_c \, .\label{mf1}
\end{eqnarray}
The first term on the right-hand side is due to the driving of the system that replaces $P(E,t)$ by $P(E-v\hbox{d}t)$ after time $\hbox{d}t$. The second and third term describe the change in $P(E,t)$ due to fires. The fraction of trees that catch fire per time $\hbox{d}t$ is $v \hbox{d}t \bar s P(E_c)$. The fraction $\alpha$ of the burning sites become evenly distributed in the interval $[0,E_0]$, while the rest is evenly distributed in the interval $[E_0,E_c)$. The fraction of sites with $E\in (E_0,E_c)$ that catch fire per time $\hbox{d}t$ is given by the third term in the second equation. Since there are no correlations in the system, the probability that a site of energy $E$ becomes involved in a fire is proportional to the number of sites with this energy. 

Since we are interested in the stationary state, we can drop the time-dependence and set $\partial P/\partial t = 0$. The solution of Eqs (\ref{mf1}) is then given by
\begin{eqnarray}
P(E) &=& \alpha P(E_c) \bar s E/E_0 \, \hbox{ for } E \le E_0 \, ;\nonumber \\
P(E) &=& {\bar s (1-\alpha)\beta \over (\bar s -1)(E_c-E_0)} + \left[
\alpha P(E_c)\bar s - {\bar s (1-\alpha)\beta \over (\bar s -1)(E_c-E_0)}\right]
\exp\left[-P(E_c)(\bar s -1)(E-E_0)/ \beta\right] \nonumber \\
&&\hbox{\hskip 2cm for } E_c>E \ge E_0\,, \label{mf2}
\end{eqnarray}
with $\beta = \int_{E_0}^{E_c} P(E') \hbox{d} E'$. 
Normalization requires $\beta = 1-\alpha P(E_c)E_0\bar s/2$. We still have to find $P(E_c)$ and $\bar s$ to complete the solution. Since we are interested in the critical behavior of the model, let us focus on the situation where $\bar s$ becomes very large. The mean fire size $\bar s$ diverges as $[(1/z)-\beta]^{-1}$ when $\beta$ approaches the percolation threshold $1/z$. This allows us to replace $(\bar s -1)$ with $\bar s$, and the right-hand side of Eqs (\ref{mf2}) depends only on the product $P(E_c) \bar s$, but not on these two quantities separately. Since $\int_0^{E_0} P(E)\hbox{d}E$ approaches a constant ($ 1-1/z$) as $\bar s$ diverges, this product must become also a constant $C$ for large $\bar s$. From $\beta = 1-C\alpha E_0/2=1/z$ follows $C=2(z-1)/z\alpha E_0$. 
The solution of our mean--field equation therefore becomes (for large $\bar s$)
\begin{eqnarray}
P(E) &=& C \alpha E/E_0 \, \hbox{ for } E \le E_0\,; \nonumber \\
P(E) &=& {(1-\alpha) \over z(E_c-E_0)} + \left[
C \alpha - {(1-\alpha)\over z(E_c-E_0)}\right]
\exp\left[-Cz(E-E_0)\right]\, \hbox{ for } E \ge E_0\,. \label{mf3}
\end{eqnarray}
This solution shows that $P(E_c)$ vanishes in the limit $E_c \to \infty$, and we conclude that $\bar s$ diverges exponentially as $\exp[Cz(E_c-E_0)]$. The exponent $\gamma$ charecterizing the stretched exponential increase of $\bar s$ is $\gamma = 1$ in mean--field theory. The exponent $\tau$ characterizing the size--distribution of fires is $\tau = 1.5$ in mean--field theory, since the density of tres with $E>E_0$ is at the (mean--field) percolation threshold. (For the mean--field theory of percolation see e.g. \cite{sta92}.)

The mean--field theory for the second model is slightly different. While the solution for $E<E_0$ has the same form as before, the probability distribution for energies larger than one is obtained from the equation
\begin{equation}
 {\hbox{d} P(E) \over\hbox{d} E} = (1-\alpha){ \bar s P(E_c)\int_E^{E_c} P(E')\hbox{d}E'\over \int_{E_0}^{E_c}\int_{E'}^{E_c} P(E'')\hbox{d}E'' \hbox{d}E'} - {P(E) P(E_c)(\bar s - 1)\over \int_{E_0}^{E_c} P(E')\hbox{d}E'} \, .\label{mf4}
\end{equation}
Introducing $R(E) = \int_E^{E_c}P(E')\hbox{d}E'$, this becomes
\begin{equation}
{\hbox{d}^2R\over \hbox{d}E^2} + a {\hbox{d}R\over \hbox{d}E} + bR = 0\,,
\label{mf5}
\end{equation}
with $a = P(E_c)(\bar s - 1) / R(E_0)$ and $b = (1-\alpha) \bar s P(E_c) / \int_{E_0}^{E_c}R(E)\hbox{d}E$. Solving this equation, and calculating $P(E) = -\hbox{d}R/\hbox{d}E$ gives
\begin{equation}
P(E) = A \exp[-\lambda_1 (E-E_0)] + B \exp[-\lambda_2 (E-E_0)]\, ,
\end{equation}
with $\lambda_{1,2} = (a/2) \pm \sqrt{(a^2/4)-b}$.  The coefficients $A$ and $B$ have to be determined by the condition that $P(E)$ is continuous at $E=E_0$, and by the normalization condition $\int_0^{E_c} P(E) \hbox{d}E = 1$. As before, the $P(E)$ depends only on the product $P(E_c)\bar s$ for large $\bar s$, which must become a constant $C=2(z-1)/\alpha z E_0$ as $\bar s$ diverges. Again, we focus on the limit of large $\bar s$, where $a=Cz$. We want to show that $\bar s$ can only diverge in the limit $E_c \to \infty$, but not for a finite $E_c = E_c^*$. For this purpose, let us assume a finite $E_c^*$ and show that this leads to contradictions. The conditions $P(E_c) = 0$ at $E_c=E_c^*$ and $P(E_0) = C\alpha$ lead to
\begin{displaymath}
P(E) = C\alpha \exp[-a(E-E_0)/2]{\sinh[(E_c-E)\sqrt{(a^2/4)-b}]\over \sinh[(E_c-E_0)\sqrt{(a^2/4)-b}]}\, \hbox{ for } E_c = E_c^* \, .
\end{displaymath}
It is easy to see that the normalization condition is $\int_{E_0}^{E_c} P(E) \hbox{d}E = 1/z$ cannot be satisfied. We have
\begin{displaymath}
\int_{E_0}^{E_c} P(E) \hbox{d}E < \alpha C / a = \alpha/z < 1/z\,,
\end{displaymath}
since $\alpha < 1$.

Consequently, $\bar s$ can only diverge in the limit $E_c \to \infty$. In this limit, $\int_{E_0}^{E_c}R(E)\hbox{d}E$ diverges as fast as $E_c - E_0$ since most sites have energies much smaller than $E_c$. From the normalization and continuity of $P(E)$ we find then to leading order in $1/(E_c-E_0)$: $A=\alpha C - B$, $B = (1-\alpha)/zQ(E_c-E_0)$, $\lambda_1 = Cz-\lambda_2$, and $\lambda_2 = 1/Q(E_c-E_0)$. The constant $Q$ depends on $\alpha$ and is defined by the equation 
$$
3Q+(1-Q)\exp(-Q^{-1}) = 1/(1-\alpha) \,.
$$
As in the first model, the asymptotic form of $P(E)$ is a simple exponential decay for $E>E_0$ with a decay constant $\lambda_1=2(z-1)/\alpha E_0$. 

\section{Conclusion}
\label{u5} 

In this paper, I have shown that certain systems with avalanche--like dynamics and two enery scales have a power--law size distribution of avalanches over many orders of magnitude when these scales are widely separated. For some of these models, a very moderate scale separation $(E_c-E_0)/E_0 \simeq 3$ gives already scaling over more than four decades. Since a separation of enery scales can occur naturally, it can give rise to power laws in nature. These models belong therefore to the class of SOC systems. They are to some degree similiar to the SOC forest--fire model. In that model, a separation of time scales for tree growth and lightning strokes is responsible for the observed power laws. When the lightning probability is very small compared to the tree growth probability, an isolated tree has to wait for a long time before it catches fire spontaneously.
 The separation of energy scales in this paper has a similiar effect: When $(E_c-E_0)/E_0$ is large, a newly grown tree (energy $E_0)$ that is not ignited by its neighbors has to wait for a long time until it reaches the threshold $E_c$ and catches fire. But there are also major differences between the two models. The models of this paper have a continuous energy scale, and a finite fraction of all burning trees remain trees after a fire. The exponent $\tau$ characterizing the size distribution of fires seems to be different.

Therefore, the models studied in this paper are a new class of SOC systems. Many variations of these models are expected to be also SOC, as long as trees with an energy close to $E_c$ have a nonvanishing probability of loosing only a small fraction of their energy during a fire. However, if burning trees turn always to empty sites, the system becomes completely synchronized when $E_c > 2E_0$. SOC systems seem to be in some cases systems that fail to synchronize, as also observed in \cite{mid95} for the earthquake model. 

Let us conclude with a caveat: Not all models that satisfy the above--mentioned critieria are SOC. To illustrate this, let us consider a model where a burning tree reduces its energy by some small amount $\Delta E \ll E_0$ with some small  probability $g$, and becomes an empty site with zero energy with probability $1-g$. 
If we begin with a completely synchronized system,  after the first fire a fraction $1-g$ of all sites will have zero energy, and the remaining sites are very close to the threshold. All these sites will burn down before the other sites reach the energy $E_0$, and there will again be a fire that spreads over the entire system. Thus, the conclusion of this paper illustrates what is already said in the introduction: Slowly driven non--equilibrium systems with avalanche--like dynamics show a richness and complexity of behavior that still has to be fully explored.

\acknowledgements
This work was supported by the Deutsche 
Forschungsgemeinschaft (DFG) under Contract No. Dr 300/1-1, and by the
NSF grant No. DMR-93-03667.

\end{document}